\documentclass[times,final]{elsarticle}
\journal{ Computational and Theoretical Chemistry}

\begin{document}

\begin{frontmatter}

\title{Schr\"odinger equation with Pauli-Fierz Hamiltonian and double well potential as model of vibrationally enhanced tunneling for proton transfer in hydrogen bond}
\author{A.E. Sitnitsky}
\ead{sitnitsky@kibb.knc.ru}

\address{Kazan Institute of Biochemistry and Biophysics, FRC Kazan Scientific Center of RAS, P.O.B. 30,
420111, Russian Federation.}

\begin{abstract}
A solution of the two-dimensional Schr\"odinger equation with Pauli-Fierz Hamiltonian and trigonometric double-well potential is obtained within the framework of the first-order of adiabatic approximation. The case of vibrational strong coupling is considered which is pertinent for polariton chemistry and (presumably) for enzymatic hydrogen transfer. We exemplify the application of the solution by calculating the proton transfer rate constant in the hydrogen bond of the Zundel ion ${\rm{H_5O_2^{+}}}$ (oxonium hydrate) within the framework of the Weiner's theory. An analytic formula is derived which provides the calculation of the proton transfer rate with the help of elements implemented in {\sl {Mathematica}}. The parameters of the model for the Zundel ion are extracted from the literature data on IR spectroscopy and quantum chemical calculations. The approach yields a vivid manifestation of the phenomenon of vibrationally enhanced tunneling, i.e., a sharp bell-shaped peak of the rate enhancement by the external vibration at its symmetric coupling to the proton coordinate. The results obtained testify that the effect of resonant activation in our model is robust and stable to variations in the types of the quadratically coupled mode (vibrational strong coupling or symmetric one).
\end{abstract}
\end{frontmatter}

\section{Introduction}
Vibrationally enhanced tunneling at proton transfer (PT) is an interdisciplinary phenomenon which can take place in various fields of chemical physics and biophysics (e.g., polariton chemistry, enzymatic hydrogen transfer (EHT), etc). In some contexts instead of "enhanced" the terms "assisted" \cite{Ahm25} or "promoted" are used. Vibrational polariton chemistry (see, e.g., [2-19] and refs. therein) is a relatively new field (some authors count its birth from 2016 \cite{Wan21}) which is actively developed at present. On the one hand this approach is deeply rooted in enormous amount of information obtained within the framework of traditional exploration of molecules with the help of IR lasers (see, e.g., \cite{May11} and refs. therein). On the other hand introducing an optical cavity (placing the molecules under study within a Fabri-P\'erot resonator) stipulates many peculiarities of such systems. Using the IR electromagnetic field of the optical cavity provides an efficient way to interfere into the internal degrees of freedom of the molecules and dynamical processes in them. The phenomenon of resonant activation (or, in contrast, suppression) of reaction rates is widely discussed for modifying chemical kinetics by optical cavities. The resonance, i.e., maximal cavity induced enhancement of the reaction rate under the vibrational resonance condition is produced in this case by mixing the electromagnetic field with quantum states of molecular systems. The field in the cavity is equivalent to a harmonic oscillator of a given frequency coupled to the molecular system. This frequency is an experimentally controllable parameter that provides wide possibilities for extremely precise influencing the molecule degrees of freedom. They are usually considered as coupled to the field oscillator in the electric dipole approximation of light-matter interaction. The theoretical description of combined quantum states of the molecular system plus IR electromagnetic field (named polaritons) is reduced to the Schr\"odinger equation (SE) with the Pauli-Fierz Hamiltonian (PFH). This combination is one of the main theoretical tools in the vibrational polariton chemistry [2-7]. In this regard constructing reliable solutions of SE with PFH is of interest for perspectives of various applications.

A conceptually related phenomenon may take place at EHT. The search for mechanisms of enzyme action has a long history but is still in active progress. Enzymes can accelerate corresponding chemical reactions by many orders of magnitude (the most efficient species do it up to factors more than $10^{20}$ \cite{Sch09}, e.g., $10^{21}$ for phosphohydrolases or $10^{26}$ for sulfate monoesters). There is a concept of dynamic mechanisms participating in reaction acceleration by enzymes (conventionally named "rate-promoting vibration" \cite{Ant01} or "gating" \cite{Bru92}, \cite{Jev17}). It is based on the growing evidence of various dynamic modes in protein scaffold of an enzyme. Arguably the birth of the concept may be traced back to \cite{War78} where electrostatic fluctuations in an enzyme active site were introduced.  Since then the possible dynamic aspect of fluctuations in the mechanism of reaction acceleration was considered in several articles (see, e.g., [23-36] and refs. therein). A particular case of such dynamic activation takes place in the mechanism for EHT [24,25,27,28,36]. It is based on the phenomenon of vibrationally enhanced tunneling at PT in hydrogen bonds (HB). It means the effect of resonant acceleration of the reaction by a coupled oscillation in some narrow frequency range \cite{Sok92}, \cite{Ham95}. For enzymes carrying out EHT there are few estimates of reaction acceleration (e.g., $10^{4}$ for alcohol dehydrogenase and $10^{12}$ for methylmalonylCoA mutase) because it is difficult to model uncatalyzed reactions for them \cite{Sch09}. There is a characteristic feature of such enzymes. Experiments show that EHT exhibits a pronounced temperature dependence of the rate constant \cite{Sch09}. Thus EHT should be treated as thermally activated tunneling.

The above mentioned conceptual proximity of such seemingly unrelated fields as polariton chemistry and EHT is based on the fact that SE with PFH has the same mathematical structure as that used for PT coupled to the heavy atoms stretching mode in HB [37-46]. Besides theoretical speculations on the similarity of the mathematical description in both fields there are optical cavity experiments testifying that the so-called vibrational strong coupling (VSC) modifies enzymes activity via water. It can lead either to the suppression of the corresponding reaction \cite{Ver19} or to the increase of the reaction rate constant \cite{Lat21}. VSC was designed to describe the electric dipole approximation of light-matter interaction in the vibrational polariton chemistry (see [2-7] and refs. therein). Most investigations in this field are devoted to the processes of molecule dissociation that require single-well potentials. Among them the famous Morse potential is mainly used in theoretical analysis due to its convenience and owing to the fact that SE with it has an exact analytic solution via associated Laguerre polynomials. However barrier crossing adiabatic chemical reactions are also of considerable interest \cite{Man22}, \cite{Li21a}, \cite{Man20}, \cite{Gam22}. In particular PT in HB is a pattern process of barrier crossing (see, e.g., [38-47] and refs. therein). The description and interpretation of barrier crossing reactions require the solution of SE with  a double-well potential (DWP) which is a one-dimensional cross-section of the potential energy surface. In DWP quantum tunneling leads to the formation of doublets (pairs of  energy levels).  Unfortunately the simplest polynomial 2-4 DWP (see, e.g., [48-51]) is not amenable to exact analytic treatment. On the other hand numerical solutions can be very accurate but provide the results in the form of tables of numbers for wave functions or energy levels at each specific set of initial conditions. It is inconvenient if such results are not a final goal but an intermediate step for further analytic work. The famous approximate analytic approach named the quasi-classical (WKB) method is known to be inaccurate for energy levels near the barrier top. For instance the comparison of various WKB expressions with the exact solution available for the discussed below trigonometric DWP (TDWP) clearly exhibits their inaccuracy for the cases when the ground state doublet is near the barrier top \cite{Sit18}. The "poorness" of WKB wave functions motivates the intoduction of the Modified Airy Function approximation which alleviates the deficiencies of those from the quasi-classical one \cite{Mar25}. Exact solutions for piecewise DWPs in terms of confluent hypergeometric functions (as, e.g., in the Quantum Mechanics textbook by Merzbacher) are of use for pedagogical purposes. However at practical applications the artificial construction of such DWPs casts some doubt upon the possible error introduced by sewing their parts. In this regard the problem of constructing practically convenient (in particular smooth, i.e., not piecewise) DWPs for which SE would have an exact analytic solution was rather pressing. As a matter of fact up to 2010-s there were no such DWPs. Since then the situation changed drastically at first with the invention of exactly solvable hyperbolic DWPs which are finite at the boundaries of the spatial variable interval (finding their applications in condensed matter physics). Then exact analytic solutions were obtained for DWPs (both hyperbolic and trigonometric) taking infinite values at the boundaries of the spatial variable interval and suitable for chemical problems. At present a number of DWPs is known [52-62] for which such solutions of SE are feasible via the confluent Heun's function (CHF) implemented in {\sl {Maple}} [54-62] or the spheroidal function (SF) [52,53,55] implemented in {\sl {Mathematica}} along with its spectrum of eigenvalues. Among them TDWP which enables one to solve SE via both functions is of particular use for applications [52,54,55,63-68] because the above mentioned option for the spectrum of eigenvalues of SF extremely facilitates the calculation of the corresponding energy levels. In contrast for DWPs solved via CHF obtaining energy levels is a challenging problem though it should be mentioned that a fine numerical algorithm was invented to overcome it (see [56]-[62] and refs. therein). As a matter of fact TDWP was known long ago. In the monograph on SF \cite{Kom76} the equation (1.9) (which actually had the form of SE with TDWP) was presented along with its solution via this special function. However explicitly this fact was recognized much later in \cite{Sch16} where many mathematical aspects of TDWP were considered but no physical consequences or implications were discussed. At the same time the solution of SE with TDWP via CHF was obtained \cite{Sit17} and its relationship to that via SF was pointed out \cite{Sit171}. It is rather notable that the results of the above mentioned activity on exactly solvable DWPs (see [52-68] and refs. therein) seem to remain poorly known to wide audience. Indeed assertions like "Unfortunately, the analytical solution of the Schr\"odinger equation is only possible for simple potentials, therefore ... " \cite{Fer21} are very typical and continue to appear up to now as a reason for the necessity to apply WKB to DWPs.

The aim of the present article is to show that TDWP enables one to obtain an approximate analytic solution (within the framework of the standard adiabatic approximation) of the two-dimensional SE with PFH. There is the expert point of view that "analytical solutions on a two-dimensional double well potential ... are still lacking in the literature" (see Ref.56 in \cite{Cha22}). We as well are not able to provide an exact analytic solution of the problem but our approximation seems to be a reliable working tool.
As an example of the application of this formalism we consider the reaction of PT in HB. For calculating PT rate we apply the Weiner's theory \cite{Wei78}, \cite{Wei78a} (see Appendix 2). We conceptually follow the line of \cite{Sit23} but there are two marked distinctions. First, in contrast to mode couplings (symmetric, anti-symmetric and squeezed ones) discussed in \cite{Sit23} we now consider VSC. The main feature of VSC is that the strength of interaction is proportional to the square root of the frequency of the oscillator \cite{Yang21}, \cite{Tri20} rather than to be a constant as for the above mentioned cases. As a result VSC takes into account a physical requirement that the strength of interaction should diminish with lowering the frequency. Second, we solve the two-dimensional SE not only in the zero-order of the adiabatic approximation as in \cite{Sit23} but up to the first order. Our terminology means the following. In the adiabatic approximation the total wave function is factorized into those of the proton and of the external harmonic oscillator. In the first stage it is done as a mathematical procedure which does not invove physical argumentation. The interaction of the proton degree of freedom with the oscillator makes its energy states to be interdependent of one another. There occurs a ladder-type system of coupled equations in which the wave function of an oscillator state depends on those of all other states. This system is exact (i.e., derived without any approximation) but it can not be stringently solved. Thus we solve it by the perturbation theory approach. The zero-order approximation physically means that initially we neglect the above mentioned interdependence of the oscillator energy states. Then in the first-order approximation we take it into account as a perturbation. In the second stage the physical argument is involved that in our bound state problem the characteristic periods of different motions are highly desparate. The frequency of the harmonic oscillator is assumed to be higher than the inverse characteristic time of the proton motion in the double well by orders of magnitude. As a result the more rapidly moving harmonic oscillator is assumed to adjust adiabatically to the slower motion of the proton along the reaction coordinate. This fact is used at the averaging of the two-dimensional SE over the variable of the harmonic oscillator coordinate. Such procedure enables us to transform this equation into the one-dimensional form which is necessary for applying the Weiner's theory. Despite all these changes and improvements we obtain a vivid manifestation of the vibrationally enhanced tunneling in the reaction rate constant similar to that of \cite{Sit23}.  We consider PT rate for intermolecular HB in the Zundel ion ${\rm{H_5O_2^{+}}}$ (oxonium hydrate ${\rm{H_2O\cdot\cdot\cdot H \cdot\cdot\cdot OH_2}}$) with the distances between oxygen atoms $R_{OO}=3.0\ \AA$ and $R_{OO}=2.8\ \AA$. On the one hand both cases provide sufficiently high barriers to exclude the contribution of the over-barrier transition into PT rate constant even at high temperature. On the other hand such high barriers make these cases to be a model for EHT. Besides for the Zundel ion the detailed data of IR spectroscopy [39-45] along with the quantum chemical {\it ab initio} calculations \cite{Yu16}, \cite{Xu18} are available. It enables us to make use of literature data for the one-dimensional cross-section of the potential energy surface from quantum-chemical calculations and model it by a suitable phenomenological TDWP.

The paper is organized as follows. In Sec.2 PFH is discussed. In Sec.3 the two-dimensional SE with PFH is solved in the adiabatic approximation. In Sec.4 we derive the expression for PT rate constant with the help of the Weiner's theory. In Sec.5 the results are discussed and the conclusions are summarized. In Appendix 1 the transformation of dimensional PFH into its dimensionless form is discussed. In Appendix 2 some formulas of the Weiner's theory are presented for the convenience of readers. In Appendix 3 the list of abbreviations used is presented.

\section{Schr\"odinger equation with Pauli-Fierz Hamiltonian}
\label{sec1}
PFH is often analyzed in its full form including the so-called dipole self-energy term \cite{Yang21}, \cite{Tri20}. However the dominant contribution is taken into account in the reduced form omitting this term (see (1) in \cite{Tri20}) and in the present article we restrict ourselves by this case. In the dimensionless form (see Appendix 1 for details) PFH of the cavity quantum electrodynamics with VSC is
\begin{equation}
\label{eq1}  H=-\frac{\partial^2}{\partial x^2}+U(x)-\delta \frac{\partial ^2 }{\partial z^2}+\frac{1}{2}\omega^2 z^2+\tilde\alpha \sqrt \omega z d(x)
\end{equation}
Further we identify the normal mode under consideration with that of the proton in HB. Thus in (\ref{eq1}) $x$ is the proton coordinate, $z$ is that of the oscillator of frequency $\omega$ produced, e.g., by the cavity electromagnetic field, $d(x)$ is the dipole moment, $\delta$ is the ratio of the reduced mass for the proton to that of the oscillator and $\tilde\alpha$ is the coupling constant, i.e., the parameter characterizing the strength of VCS. For the one-dimensional cross-section of the potential energy surface $U(x)$ we make use of TDWP \cite{Sit17}, \cite{Sit18}
\begin{equation}
\label{eq2} U(x)=\left(m^2-\frac{1}{4}\right)\ \tan^2 x-p^2\sin^2 x
\end{equation}
Here $-\pi/2 \leq x \leq \pi/2$, $m$ is an integer number and $p$ is a real number. The two parameters of TDWP $m$ and $p$ are related to two main characteristics of the potential energy surface, i.e., the barrier hight and the barrier width (see Appendix 1). The examples of TDWP for intermolecular HB in the Zundel ion are presented in Fig.1 and Fig.4.
\begin{figure}
\begin{center}
\includegraphics* [width=\textwidth] {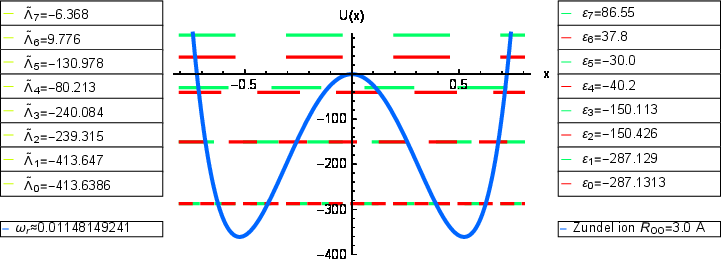}
\end{center}
\caption{The trigonometric double-well potential (\ref{eq2}) at the values of the parameters $m=57$; $p=76$. The parameters are chosen to describe the hydrogen bond in the Zundel ion ${\rm{H_5O_2^{+}}}$ (oxonium hydrate) for the case $R_{OO}=3.0\ \AA$ (they are extracted from the data of quantum chemistry \cite{Yu16}, \cite{Xu18}). At the right of Fig.1 the energy levels in the absence of an external oscillation (given by (\ref{eq7})) are presented. The energy levels of polaritonic states (given by (\ref{eq36})) are presented (but not ploted) to the left of Fig.1. They are calculated for the coupling constant $\alpha=5$ and at the frequency of the external oscillator $\omega_r=0.01148149240963448$ corresponding to the peak in Fig.2.} \label{Fig.1}
\end{figure}
As a result the dimensionless form of the stationary two-dimensional SE (see Appendix 1)
\begin{equation}
\label{eq3} H \Phi(x,z)=\Lambda \Phi(x,z)
\end{equation}
takes the form
\begin{equation}
\label{eq4}\Biggl\{\delta \frac{\partial ^2 }{\partial z^2}+\frac{\partial^2}{\partial x^2}+\Lambda-\left(m^2-\frac{1}{4}\right)\ \tan^2 x+p^2\sin^2 x-\frac{1}{2}\omega^2 z^2-\tilde\alpha \sqrt \omega z d(x)\Biggr \}\Phi(x,z)=0
\end{equation}
\section{Solution of two-dimensional Schr\"odinger equation in adiabatic approximation}
The exact analytic solution of one-dimensional SE with TDWP (\ref{eq2}) for the energy levels $\epsilon_q$ and the corresponding wave functions $\psi_q (x)$
\begin{equation}
\label{eq5}  \psi''_q(x)+\left[\epsilon_q-\left(m^2-\frac{1}{4}\right)\ \tan^2 x+p^2\sin^2 x\right]\psi_q(x)=0
\end{equation}
is available \cite{Sit18}
\begin{equation}
\label{eq6} \psi_q (x)=\sqrt {\cos x}\ \bar S_{m(q+m)}\left(p;\sin x\right)
\end{equation}
where $q=0,1,2,...$ and $\bar S_{m(q+m)}\left(p;s\right)$ is the normalized angular prolate spheroidal function \cite{Kom76}. It is implemented in {\sl {Mathematica}}
as $\rm{SpheroidalPS}[(q+m),m,ip,s]$ where $i=\sqrt {-1}$ (note that the representation in this program software package is a non-normalized one). The energy levels are
\begin{equation}
\label{eq7} \epsilon_q=\lambda_{m(q+m)}\left(p\right)+\frac{1}{2}-m^2-p^2
\end{equation}
Here $\lambda_{m(q+m)}\left(p\right)$ is the spectrum of eigenvalues for $\bar S_{m(q+m)}\left(p;s\right)$. It is implemented in {\sl {Mathematica}} as $\lambda_{m(q+m)}\left(p\right)\equiv \rm{SpheroidalEigenvalue}[(q+m),m,ip]$. For TDWP the position of the right minimum is defined by the requirement
\begin{equation}
\label{eq8} \cos\ x_{min}=\left[\frac{m^2-1/4}{p^2}\right]^{1/4}
\end{equation}

The functions $\psi_q (x)$ form a full orthogonal system so that within the framework of the adiabatic approximation we seek the solution of (\ref{eq4}) as an expansion over it (see, e.g., (1.3) in \cite{Dav71})
\begin{equation}
\label{eq9} \Phi(x,z)=\sum_{q=0}^{\infty}\ \varphi^q(z) \psi_q (x)
\end{equation}
For the coefficients $\varphi^q(z)$ we obtain a closed and self-consistent system of differential equations
\begin{equation}
\label{eq10} \left[\delta \frac{d^2}{dz^2}+\Lambda_q-\epsilon_q-\frac{\omega^2 z^2}{2}-\tilde\alpha \sqrt \omega a_{qq} z\right]\varphi^q(z)=\tilde\alpha \sqrt \omega z \sum_{l=0, l \neq q}^{\infty}\ a_{ql}\varphi^l(z)
\end{equation}
where we denote
\begin{equation}
\label{eq11} a_{ql}=\int_{-\pi/2}^{\pi/2}dx\  d(x) \psi_q(x)\psi_l(x)
\end{equation}
It noteworthy that the system of equations (\ref{eq10}) is an exact one. Then our aim is to solve it up to the first-order approximation.
In the zero-order approximation we neglect the non-diagonal terms in the right-hand side of (\ref{eq10}) and obtain an equation for $\ ^{(0)}\varphi^q(z)$
\begin{equation}
\label{eq12}  \left[\delta \frac{d^2}{dz^2}+\ ^{(0)}\Lambda_q-\epsilon_q+\frac{\tilde\alpha^2 a_{qq}^2}{2\omega}-\frac{\omega^2}{2}\left(z+\frac{\tilde\alpha a_{qq}}{\omega^{3/2}}\right)^2\right]\ ^{(0)}\varphi^q(z)=0
\end{equation}
Its solution is possible if the states of the oscillator are quantized with the corresponding quantum number $j=0,1,2,...$ and energy levels are
\begin{equation}
\label{eq13} \ ^{(0)}\Lambda^j_q=\lambda_{m(q+m)}\left(p\right)+\frac{1}{2}-m^2-p^2-\frac{\tilde\alpha^2 a_{qq}^2}{2\omega}+(4j+1)\omega \sqrt {\frac{\delta}{2}}
\end{equation}
The solution has the form
\begin{equation}
\label{eq14} \ ^{(0)}\varphi^q_j(z)=\ ^{(0)}A\exp\left[-\frac{\omega}{2\sqrt{2\delta}}\left(z+\frac{\tilde\alpha a_{qq}}{\omega^{3/2}}\right)^2\right]\frac{j!(-2)^j}{(2j)!} H_{2j}\left(\left(\frac{2\omega^2}{\delta}\right)^{1/4}\left(z+\frac{\tilde\alpha a_{qq}}{\omega^{3/2}}\right)\right)
\end{equation}
where $H_n(x)$ is the Hermite polynomial and $\ ^{(0)}A$ is the normalization constant. It can be obtained with the help of N2.20.16.6 from \cite{Pr03}
\begin{equation}
\label{eq15} \ ^{(0)}A^{-2}=\left[\frac{j!(-2)^j}{(2j)!}\right]^2 2^{4j}\sqrt{\frac{\sqrt{2\delta}}{\omega}}\Gamma \left(2j+\frac{1}{2}\right)\ _2F_1\left(-2j,-2j,\frac{1}{2}-2j;-\frac{1}{2}\right)
\end{equation}
where $\ _2F_1\left(a,b,c;x\right)$ is the hypergeometric function and $\Gamma (x)$ is the gamma-function.
With the help of (\ref{eq14}) we calculate the average value of $z$ in the zero-order approximation
\begin{equation}
\label{eq16} \ ^{(0)}<z>_q\ =\int_{-\infty}^{\infty}dz\ z\  \left[\ ^{(0)}\varphi^q_j(z)\right]^2=-\frac{\tilde\alpha a_{qq}}{\omega^{3/2}}
\end{equation}
In the first-order approximation we have
\begin{equation}
\label{eq17}\left[\delta \frac{d^2}{dz^2}+\ ^{(1)}\Lambda_q^j-\epsilon_q-\frac{\omega^2 z^2}{2}-\tilde\alpha \sqrt \omega a_{qq} z\right]\ ^{(1)}\varphi^q_j(z)= \tilde\alpha \sqrt \omega z \sum_{l=0, l \neq q}^{\infty}\ a_{ql}\ ^{(0)}\varphi^l_j(z)
\end{equation}
We make the transformation under the sum in the right-hand side
\begin{equation}
\label{eq18} \ ^{(0)}\varphi^l_j(z)=\ ^{(0)}\varphi^l_j(z)\frac{\ ^{(1)}\varphi^q_j(z)}{\ ^{(1)}\varphi^q_j(z)}=\ ^{(1)}\varphi^q_j(z)\frac{\ ^{(0)}\varphi^l_j(z)}{\ ^{(1)}\varphi^q_j(z)}\approx\ ^{(1)}\varphi^q_j(z)\frac{\ ^{(0)}\varphi^l_j(z)}{\ ^{(0)}\varphi^q_j(z)}\approx\ ^{(1)}\varphi^q_j(z)
\frac{\ ^{(0)}\varphi^l_j\left(\ ^{(0)}<z>_q\right)}{\ ^{(0)}\varphi^q_j\left(\ ^{(0)}<z>_q\right)}=\ ^{(1)}\varphi^q_j(z)h_j^{ql}
\end{equation}
where we denote
\begin{equation}
\label{eq19}h_j^{ql}=\frac{H_{2j}\left(\left(\frac{2}{\delta}\right)^{1/4}\frac{\tilde\alpha}{\omega}\left(a_{ll}-a_{qq}\right)\right)}
{H_{2j}(0)}\exp\left[-\frac{\tilde\alpha^2}{2\sqrt{2\delta}\omega^2}\left(a_{ll}-a_{qq}\right)^2\right]
\end{equation}
The last step in (\ref{eq18}) $z \longrightarrow \ ^{(0)}<z>_q$ is justified by the fact that in (\ref{eq14}) only the region $z \simeq \ ^{(0)}<z>_q$ is of importance due to the exponential factor. Its noteworthy that $H_{2j}(0)\not= 0$ at any $j$ so that there can not be infinity because of the denominator in (\ref{eq19}).

Here it is necessary to stress that to make the approach of practical use one should somehow get rid of the sum over $j$ in the final expression for the reaction rate constant. Otherwise the computation with two sums (over $q$ and $j$) is unreasonably cumbersome. To eliminate the sum over $j$ is possible in particular if the numerator and the denominator (which will be a partition function) of the expression for the rate constant are completely factorized as a product of independent sums over $j$ and $q$. To make it feasible we need $h_j^{ql}$ to be $j$-independent. For this reason a further approximation is necessary
\begin{equation}
\label{eq20}h_j^{ql}\approx h_q^l
\end{equation}
where
\begin{equation}
\label{eq21}h_q^l=\exp\left[-\frac{\tilde\alpha^2}{2\sqrt{2\delta}\omega^2}\left(a_{ll}-a_{qq}\right)^2\right]
\end{equation}
Such dropping out the ratio $H_{2j}\left(\left(\frac{2}{\delta}\right)^{1/4}\frac{\tilde\alpha}{\omega}\left(a_{ll}-a_{qq}\right)\right)/H_{2j}(0)$ in (\ref{eq19}) may be justified as follows. One can see from the exponential factor that only the case when $a_{ll}-a_{qq} \rightarrow 0$ is of importance so that this ratio $\rightarrow 1$. Thus the ratio can be dropped out in this case. In the opposite case (when the value $a_{ll}-a_{qq}\not\simeq 0$ makes this ratio to differ from $1$ substantially) the exponential factor is small. This circumstance actually makes the corresponding terms in the right-hand side of (\ref{eq17}) to be of no importance so that braking (\ref{eq20}) is of no significance in this case.

As a result we have an approximate equation
\begin{equation}
\label{eq22} \Biggl\{\delta \frac{d^2}{dz^2}+\ ^{(1)}\Lambda_q^j-\epsilon_q+\frac{\tilde\alpha^2}{2\omega}\left(a_{qq}+\sum_{l=0, l \neq q}^{\infty} a_{ql}h_q^l\right)^2-\frac{\omega^2}{2}\left[z+\frac{\tilde\alpha }{\omega^{3/2}}\left(a_{qq}+\sum_{l=0, l \neq q}^{\infty}a_{ql}h_q^l\right)\right]^2\Biggr\}\ ^{(1)}\varphi^q_j(z)=0
\end{equation}
One can see that in our approach the first-order approximation leads to changing the value which in the zero-order is merely $a_{qq}$. The solution of (\ref{eq22}) is
\[
\ ^{(1)}\varphi^q_j(z)=\ ^{(1)}A\exp\left\{-\frac{\omega}{2\sqrt{2\delta}}\left[z+\frac{\tilde\alpha }{\omega^{3/2}}\left(a_{qq}+\sum_{l=0, l \neq q}^{\infty} a_{ql}h_q^l\right)\right]^2\right\}\times
\]
\begin{equation}
\label{eq23} \frac{j!(-2)^j}{(2j)!} H_{2j}\left\{\left(\frac{2\omega^2}{\delta}\right)^{1/4}\left[z+\frac{\tilde\alpha }{\omega^{3/2}}\left(a_{qq}+\sum_{l=0, l \neq q}^{\infty}a_{ql}h_q^l\right)\right]\right\}
\end{equation}
where $\ ^{(1)}A$ is the normalization constant. One can easily verify that $\ ^{(1)}A=\ ^{(0)}A$ where $\ ^{(0)}A$ is given by (\ref{eq15}).
For the energy levels we have
\begin{equation}
\label{eq24}\ ^{(1)}\Lambda^j_q=\lambda_{m(q+m)}\left(p\right)+\frac{1}{2}-m^2-p^2- \frac{\tilde\alpha^2}{2\omega}\left(a_{qq}+\sum_{l=0, l \neq q}^{\infty} a_{ql}h_q^l\right)^2+(4j+1)\omega \sqrt {\frac{\delta}{2}}
\end{equation}

To move further we should specify the dipole moment $d(x)$. The shape of the electric dipole function in the polariton chemistry is thoroughly discussed in \cite{Tri20}. A universal phenomenological form via an exponential function is suggested  but for our purpose the expansion up to second order is more convenient because it clearly reveals the contributions from the symmetric and anti-symmetric parts. Such expansion is usually sufficient to describe various cases that can be encountered in the vibrational polariton chemistry \cite{Tri20}. In our case of the Zundel ion there is no constant contribution and the equilibrium position of the dipole moment is at $x=0$ so that the expansion is $d(x)=C_1 x+ C_2 x^2$. This form takes into account the commonly accepted fact that the dipole moment deflects to slower growth than the linear one (see, e.g., Fig.1 in \cite{Li21a} or Fig. 10.54 in \cite{Atk09}).
For TDWP defined at $-\pi/2 \leq x \leq \pi/2$ it is more natural for the mathematical convenience to make for $x$ in the dipole moment the transformation $x \longrightarrow \sin x$ so that
\begin{equation}
\label{eq25} d(x)=C_1 \sin x+ C_2 \sin^2 x
\end{equation}
Such transformation is logically consistent to the above mentioned deflection while its degree is regulated by the constants $C_1$ and $C_2$. We denote
\[
c_q(\omega)=\tilde\alpha\left(a_{qq}+\sum_{l=0, l \neq q}^{\infty} a_{ql}h_q^l\right)=\tilde\alpha \Biggl[\int_{-\pi/2}^{\pi/2}dx\  d(x) \left(\psi_{q}(x)\right)^2+\sum_{l=0, l \neq q}^{\infty} h_q^l\int_{-\pi/2}^{\pi/2}dx\  d(x) \psi_q(x)\psi_l(x)\Biggr]=
\]
\begin{equation}
\label{eq26}\tilde\alpha \Biggl\{C_2 \int_{-\pi/2}^{\pi/2}dx\  \sin^2 x \left(\psi_{q}(x)\right)^2+ \sum_{l=0, l \neq q}^{\infty}h_q^l\int_{-\pi/2}^{\pi/2}dx\  \left[C_1 \sin x+ C_2 \sin^2 x\right] \psi_q(x)\psi_l(x)\Biggr\}
\end{equation}
Here we take into account that the odd $\sin x$ term produces zero contribution in combination with the even function $\left(\psi_{q}(x)\right)^2$. We denote
\begin{equation}
\label{eq27} \alpha=C_2\tilde \alpha;\ \ \ \ \ \ \ \ \ \ \ \ \ \ \ \ \ \ \ \ \ \ \ \ \ \ \ \ \ \ \ \ \ \ \ \ \ \gamma=C_1\tilde \alpha
\end{equation}
Making use of (\ref{eq6}) we have
\[
 c_q(\omega)=\alpha\int_{-1}^{1}d\eta\ \eta^2\ \left[\bar S_{m(q+m)}\left(p;\eta\right)\right]^2+\sum_{l=0, l \neq q}^{\infty}\exp\Biggl\{-\frac{\alpha^2}{2\sqrt{2\delta}\omega^2}\Biggl(\int_{-1}^{1}d\eta\ \eta^2\ \left[\bar S_{m(l+m)}\left(p;\eta\right)\right]^2-
\]
\begin{equation}
\label{eq28}\int_{-1}^{1}d\eta\ \eta^2\ \left[\bar S_{m(q+m)}\left(p;\eta\right)\right]^2\Biggr)^2\Biggr\}\int_{-1}^{1}d\eta\ \left[\gamma \eta+ \alpha\eta^2 \right]\bar S_{m(q+m)}\left(p;\eta\right)\bar S_{m(l+m)}\left(p;\eta\right)
\end{equation}
In these designations we have
\[
\ ^{(1)}\varphi^q_j(z)=\left[2^{4j}\sqrt{\frac{\sqrt{2\delta}}{\omega}}\Gamma \left(2j+\frac{1}{2}\right)\ _2F_1\left(-2j,-2j,\frac{1}{2}-2j;-\frac{1}{2}\right)\right]^{-1/2}\times
\]
\begin{equation}
\label{eq29}\exp\left\{-\frac{\omega}{2\sqrt{2\delta}}\left[z+\frac{c_q(\omega) }{\omega^{3/2}}\right]^2\right\} H_{2j}\left\{\left(\frac{2\omega^2}{\delta}\right)^{1/4}\left[z+\frac{c_q(\omega)}{\omega^{3/2}}\right]\right\}
\end{equation}
\begin{equation}
\label{eq30} \ ^{(1)}\Lambda^j_q=\lambda_{m(q+m)}\left(p\right)+\frac{1}{2}-m^2-p^2-\frac{\left(c_q(\omega)\right)^2}{2\omega}+(4j+1)\omega \sqrt {\frac{\delta}{2}}
\end{equation}
With the help of (\ref{eq29}) we calculate the average value of $z$ in the first-order approximation
\begin{equation}
\label{eq31} \bar z_q\buildrel\rm def\over =\ ^{(1)}<z>_q\ =\int_{-\infty}^{\infty}dz\ z\left(\ ^{(1)}\varphi^q_j(z)\right)^2=-\frac{c_q(\omega)}{\omega^{3/2}}
\end{equation}
Summarizing the results we write out the required wave function (the solution of (\ref{eq4})) corresponding to the $q$-th state of the particle in TDWP and the $j$-th state of the oscillator
\begin{equation}
\label{eq32}\Phi(x,z,q,j)=\ ^{(1)}\varphi^q_j(z)\psi_q (x)
\end{equation}
This solves the problem of SE for PFH with TDWP in the first-order of adiabatic approximation. As an example of the application of this formula in the next Sec. we use it for calculating PT rate constant in HB under the influence of an external oscillator.

\section{Proton transfer rate constant}
We want to apply the Weiner's theory \cite{Wei78}, \cite{Wei78a} to the posed above problem. This theory has several restrictions. First, it deals with a one-dimensional SE. For this reason we average (\ref{eq4}) over the variable $z$, i.e., we change $z$ by its average value $\bar z_q$ given by (\ref{eq31})
\begin{equation}
\label{eq33}z \longrightarrow  \bar z_q=-\frac{c_q(\omega)}{\omega^{3/2}}
\end{equation}
 and drop the derivative of the wave function on $z$. Second, the Weiner's theory was developed for symmetric DWPs. The attempt to generalize it to an asymmetric case made in \cite{Wei81} unfortunately can not be directly applied to our TDWP because it was developed for a particular piecewise harmonic DWP. In the present article we remain ourselves within the framework of the symmetric case by choosing $C_1=0$ in the expression for the dipole moment (i.e., we further set $\gamma=0$). Besides it should be stressed that there is one more obstacle (of purely technical character rather than of principal one as for the above mentioned problem) that forces us to consider namely the symmetric case. The analytic solution for asymmetric one-dimensional TDWP is known via the generalized (Coulomb) SF \cite{Sit18}, \cite{Sit19} and the corresponding package for this function in {\sl {Mathematica}} was developed long ago by Falloon \cite{Fal01}. However it is unfortunately still not implemented in the standard versions of this software. Overcoming this obstacle is very troublesome at present because it requires obtaining a permission to implement the external package for generalized SF and a tiresome procedure of such implementation. Another way to overcome the problem is to resort to the solution of SE with asymmetric TDWP obtained in \cite{Sit17}, \cite{Sit171} via CHF. This special function is implemented in {\sl {Maple}} and enables one to include naturally the asymmetric case into consideration. However such way as mentioned in Introduction has its troublesome drawbacks with calculating the energy levels.

For all these reasons in the present article we restrict ourselves by the symmetric case $\gamma=0$ ($C_1=0$). As a result the averaged one-dimensional SE takes the form
\begin{equation}
\label{eq34}\Biggl\{\frac{\partial^2}{\partial x^2}+\tilde\Lambda_q-\left(m^2-\frac{1}{4}\right)\ \tan^2 x+\left(p^2+\frac{\alpha c_q(\omega)}{\omega}\right)\sin^2 x\Biggr \}\tilde\psi_q (x)=0
\end{equation}
Its exact analytic solution for the normalized wave function $\tilde\psi_q (x)$ is
\begin{equation}
\label{eq35} \tilde\psi_q (x)=\sqrt {\cos x}\ \bar S_{m(q+m)}\left(\sqrt {p^2+\frac{\alpha c_q(\omega)}{\omega}};\sin x\right)
\end{equation}
Here $q=0,1,2,...$ and $\bar S_{m(q+m)}\left(p;s\right)$ is the normalized angular SF. The energy levels are determined by the relationship
\begin{equation}
\label{eq36} \tilde\Lambda_q=\lambda_{mq}\left(\sqrt {p^2+\frac{\alpha c_q(\omega)}{\omega}}\right)+\frac{1}{2}-m^2-p^2-\frac{\alpha c_q(\omega)}{\omega}
\end{equation}
Here $\lambda_{mq}\left(p\right)$ is the spectrum of eigenvalues for the function $\bar S_{m(q+m)}\left(p;s\right)$. By analogy with the polariton chemistry we call these states for TDWP in the interaction with the mode of the external oscillator (i.e., having the mixed energy levels (\ref{eq36})) as polaritonic ones. It is notable that at frequencies where the phenomenon of vibrationally enhanced tunneling discussed below occurs (see Fig.2 and Fig.5) the energy levels of the polaritonic states $ \tilde\Lambda_q$ may have a reversed order (an energy level with $q+1$ has lower energy than that with $q$) as can be seen in Fig.1 and Fig.4. This fact took place for symmetric coupling as well \cite{Sit23}. \\

In the right-hand side well of TDWP ($0 \leq x$) there are two turning points $x_q^{(0)}$ and $x_q^{(1)}$ ($0<x_q^{(0)}<x_{min}<x_q^{(1)}<\pi/2$).
They are determined as the solutions of the equation
\begin{equation}
\label{eq37} \tilde\Lambda_q=U(x)
\end{equation}
To make use of the Weiner's approach we need to represent the wave function $\tilde\psi_q (x)$ in the quasi-classical form (see Appendix 2). We write it as the well-known WKB formula (see, e.g., \cite{Lan74})
\begin{equation}
\label{eq38} \tilde\psi_q (x)=\frac{B_q }{\sqrt {P_q (x)}}\cos\left(\int_{x_q^{(0)}}^x d\xi\ P_q (\xi) -\pi/4\right)
\end{equation}
for $ x_q^{(0)}< x <  x_q^{(1)}$ where $\tilde\Lambda_q > U(x_{min})$.  Otherwise (at  $\tilde\Lambda_q \leq U(x_{min})$ which is as well possible for polaritonic states)  the Weiner's approach is inapplicable and one should take $J_q\mid T_q\mid^2=0$ in the formula for PT rate constant. The WKB expression for $P_q(x)$  (see, e.g., \cite{Lan74}) in our notation is
\begin{equation}
\label{eq39}  P_q^{WKB}(x)=\sqrt{\tilde\Lambda_q-U(x)}
\end{equation}
For the bottom of TDWP ($x=x_{min}$) we have from (\ref{eq38}) and (\ref{eq39})
\begin{equation}
\label{eq40} \tilde\psi_q \left(x_{min}\right)=\frac{B_q }{\left(\tilde\Lambda_q-U \left(x_{min}\right)\right)^{1/4}}\cos\left(\int_{x_q^{(0)}}^{x_{min}}d\xi\  \left(\tilde\Lambda_q-U(\xi)\right)^{1/2} -\pi/4\right)
\end{equation}
Thus finally we obtain for even $q$ ($q=2n$ where $n=0,1,2,...$)
\begin{equation}
\label{eq41} B_{2n}^2= \tilde\psi_{2n}^2 \left(x_{min}\right)\left(\tilde\Lambda_{2n}-U \left(x_{min}\right)\right)^{1/2}\cos^{-2}\left(\int_{x_{2n}^{(0)}}^{x_{min}}d\xi\  \left(\tilde\Lambda_{2n}-U(\xi)\right)^{1/2} -\pi/4\right)
\end{equation}
where $\tilde\psi_{2n} (x)$ and $\tilde\Lambda_{2n}$ are given by  (\ref{eq35}) and  (\ref{eq36}) respectively.
As a result the desired product is
\[
J_{2n} \mid T_{2n} \mid^2=\Biggl[ \tilde\Lambda_{2n+1}(\omega)- \tilde\Lambda_{2n}(\omega)\Biggr]^2/B_{2n}^2=
\]
\begin{equation}
\label{eq42} \Biggl[ \tilde\Lambda_{2n+1}(\omega)- \tilde\Lambda_{2n}(\omega)\Biggr]^2\tilde\psi_{2n}^{-2} \left(x_{min}\right)\left(\tilde\Lambda_{2n}-U \left(x_{min}\right)\right)^{-1/2}\cos^{2}\left(\int_{x_{2n}^{(0)}}^{x_{min}}d\xi\  \left(\tilde\Lambda_{2n}-U(\xi)\right)^{1/2} -\pi/4\right)
\end{equation}
if $\tilde\Lambda_{2n}(\omega) >U \left(x_{min}\right)$ and
\begin{equation}
\label{eq43}J_{2n} \mid T_{2n} \mid^2=0
\end{equation}
if $\tilde\Lambda_{2n}(\omega) \leq U \left(x_{min}\right)$.
Thus we express the value $J_{2n} \mid T_{2n} \mid^2$ via the calculable values implemented in {\sl {Mathematica}}.\\

We introduce the dimensionless inverse temperature $\beta$ (see Appendix 1). Further we restrict ourselves to the case of the Boltzmann statistics. This case is certainly valid for the relatively high temperature range $200\ K \leq T \leq 400\ K$ ($0.0345 \leq \beta\leq 0.069$) where most experiments on HB are carried out or EHT takes place. It should be stressed that modern researches on HB in the Zundel ion descend to temperatures up to $T= 1\ K$ \cite{Kal09}. To extend the theory into extremely low temperature region the so-called harmonic quantum correction factor \cite{Kal09} is introduced that actually replaces the Boltzmann distribution function for the population of the energy levels by the Bose-Einstein one. However in the present article we leave aside such subtleties and make use of the Boltzmann statistics. Then the partition function is calculated with the help of the energy levels $\ ^{(1)}\Lambda^j_q$ given by the formula (\ref{eq30})
\begin{equation}
\label{eq44} Z(\beta, \omega)=\sum_{q=0}^{\infty}\sum_{j=0}^{\infty}\exp\left[-\beta \ ^{(1)}\Lambda^j_q(\omega)\right]=
\sum_{j=0}^{\infty}e^{-\beta \left[(4j+1) \omega \sqrt {\frac{\delta}{2}}+\frac{1}{2}-m^2-p^2\right]}\sum_{q=0}^{\infty}e^{-\beta \left[\lambda_{m(q+m)}\left(p\right)-\frac{\left(c_q\right)^2}{2\omega}\right]}
\end{equation}
It should be stressed that the Weiner's theory (see Appendix 2) is originally written for the infinite range of the space variable $-\infty < x < \infty$ with the requirement $\mid \psi_q (x) \mid \rightarrow 0$ at $x \rightarrow \pm \infty$. In our case of TDWP we have the requirement $\mid \psi_q (x) \mid \rightarrow 0$ at $x \rightarrow \pm \pi/2$. For this reason we apply the corresponding formulas to the case $-\pi/2 \leq x \leq \pi/2$.
The two-dimensional generalization of (\ref{eq64}) from Appendix 2 is
\begin{equation}
\label{eq45}k(\beta, \omega)\approx\frac{1}{Z(\beta, \omega)}\Biggl\{\sum_{n}\sum_{j=0}^{\infty}e^{-\beta \ ^{(1)}\Lambda^j_{2n}(\omega)}\Biggl[ \tilde\Lambda_{2n+1}(\omega)- \tilde\Lambda_{2n}(\omega)\Biggr]^2/B_{2n}^2+\sum_{l=2N+2}^{\infty}\sum_{j=0}^{\infty}e^{-\beta \ ^{(1)}\Lambda^j_l(\omega)}\Biggr\}
\end{equation}
The sum over $n=0,1,2,..., N\ $ is that over $N+1$ doublets below the barrier top. As both $B_{2n}$ and the difference $\tilde\Lambda_{2n+1}(\omega)-\ \tilde\Lambda_{2n}(\omega)$ (see (\ref{eq36})) do not depend on $j$ the numerator is factorized into the sums over $n$ or $l$ and $j$. The sum over $j$ is the same as in the expression of the partition function so that they are canceled out from the numerator and denominator. As a result we obtain for the rate constant
\[
k(\beta, \omega)\approx\Biggl \{\sum_{n=0}^N\ e^{-\beta \left(\lambda_{m(2n+m)}\left(p\right)-\frac{\left(c_{2n}(\omega)\right)^2}{2\omega}\right)}\Biggl[\left(\lambda_{m(2n+1+m)}\left(\sqrt {p^2+\frac{\alpha c_{2n+1}(\omega)}{\omega}}\right)-\frac{\alpha c_{2n+1}(\omega)}{\omega}
\right)-
\]
\[
\left(\lambda_{m(2n+m)}\left(\sqrt {p^2+\frac{\alpha c_{2n}(\omega)}{\omega}}\right)-\frac{\alpha c_{2n}(\omega)}{\omega}\right)\Biggr]^2\cos^{-1} x_{min}\ \bar S^{-2}_{m(q+m)}\left(\sqrt {p^2+\frac{\alpha c_q(\omega)}{\omega}};\sin x_{min}\right)\times
\]
\[
\left[\left(\lambda_{m(2n+m)}\left(\sqrt {p^2+\frac{\alpha c_{2n}(\omega)}{\omega}}\right)+\frac{1}{2}-m^2-p^2-\frac{\alpha c_{2n}(\omega)}{\omega}\right)-U \left(x_{min}\right)\right]^{-1/2}\times
\]
\[
\cos^{2}\left(\int_{x_{2n}^{(0)}}^{x_{min}}d\xi\  \left[\left(\lambda_{m(2n+m)}\left(\sqrt {p^2+\frac{\alpha c_{2n}(\omega)}{\omega}}\right)+\frac{1}{2}-m^2-p^2- \frac{\alpha c_{2n}(\omega)}{\omega}\right)-U \left(\xi\right)\right]^{1/2} -\pi/4\right)+
\]
\begin{equation}
\label{eq46}\sum_{l=2N+2}^{\infty}\ e^{-\beta \left(\lambda_{m(l+m)}\left(p\right)-\frac{\left(c_l(\omega)\right)^2}{2\omega}\right)}\Biggr\}\left[\sum_{q=0}^{\infty}\ e^{-\beta \left(\lambda_{m(q+m)}\left(p\right)-\frac{\left(c_q(\omega)\right)^2}{2\omega}\right)}\right]^{-1}
\end{equation}

\section{Results and discussion}
Our approach testifies that even the restricted case of purely symmetric coupling ($\gamma=0$) is able to produce rather interesting results. As in \cite{Sit23} we exemplify (\ref{eq46}) by its application to the Zundel ion with the distance between oxygen atoms $R_{OO}=3.0\ \AA$. The parameters of TDWP are $p=76$ and $m=57$. In this case there are three doublets below the barrier top (see Fig.1) that means $N=2$ in (\ref{eq46}). In the calculations of the rate constant we also take into account two levels above the barrier top (i.e., replace $\infty$ in (\ref{eq46}) by $N_{max}=7$) to make sure that the contribution of the over-barrier transition can be discarded. In all calculations we take the model value $\delta=1$ for the ratio of the mass for the proton to that of the oscillator. The value of the coupling constant $\alpha$ is chosen to be $\alpha=5$. Variations of $\{p,m,\delta,\alpha\}$ are discussed below.\\
\begin{figure}
\begin{center}
\includegraphics* [width=\textwidth] {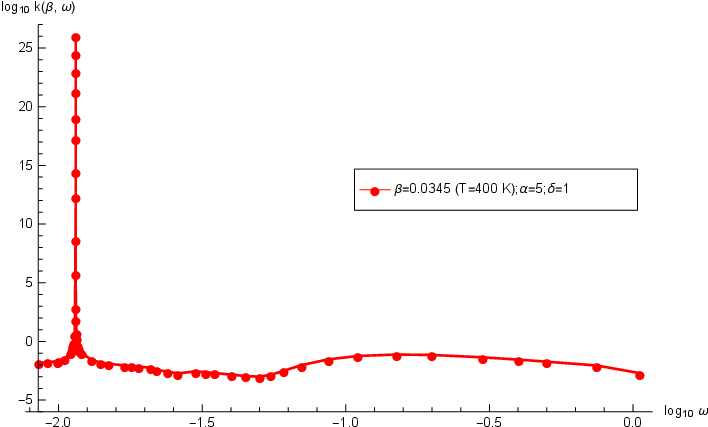}
\end{center}
\caption{The dependence of the proton transfer rate constant on the frequency of the external oscillator in the Zundel ion ${\rm{H_5O_2^{+}}}$ (oxonium hydrate) with $R_{OO}=3.0\ \AA$ (the values of the parameters for TDWP are $m=57$; $p=76$) at high temperature $T=400\ K$ ($\beta=0.0345$). The value of the coupling constant between the proton coordinate and that of the oscillator is $\alpha=5$ and the ratio of the mass for the proton to that of the oscillator is $\delta=1$. } \label{Fig.2}
\end{figure}

Quadratically coupled mode is known for a long time to be a promoting (or gating) one \cite{May11}. Our results corroborate this general conclusion. Indeed Fig.2 exhibits that the symmetric coupling of the external vibration to the proton coordinate yields vivid manifestation of resonant activation, i.e., a sharp bell-shaped peak of PT rate enhancement. Fig.2 shows that there is a high peak at the resonant frequency $\omega_r=0.01148149240963448$ providing PT rate enhancement up to $10^{26}$.  At such frequency only the second and the third doublets ($n=1$ and $n=2$) satisfy the requirement $\tilde\Lambda_{2n}(\omega) >U \left(x_{min}\right)$. Thus we set $J_{2n} \mid T_{2n} \mid^2=0$ for $n=0$ in (\ref{eq46}) at $\omega=\omega_r$. The effect of PT rate enhancement is due to resonance absorbtion of oscillator energy by the second doublet of polaritonic energy levels (see below) leading to extremely large value of $J_2 \mid T_2\mid^2$ at $\omega_r$.  Fig.3 shows that this phenomenon of vibrationally enhanced tunneling is temperature dependent.  With the decrease of temperature the height of the peak is slightly decreased while the resonance frequency remains unchanged. \\
\begin{figure}
\begin{center}
\includegraphics* [width=\textwidth] {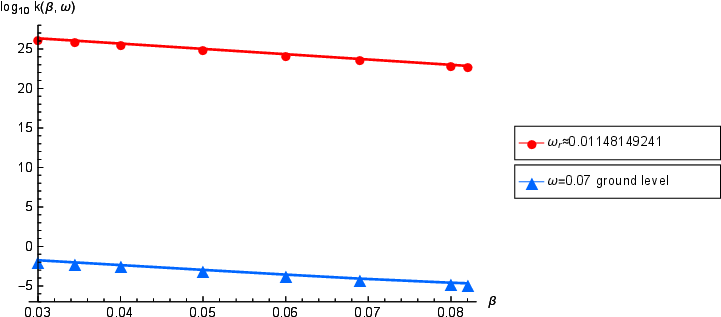}
\end{center}
\caption{The dependence of the proton transfer rate constant on the inverse temperature in the Zundel ion ${\rm{H_5O_2^{+}}}$ (oxonium hydrate) with $R_{OO}=3.0\ \AA$ at the resonance frequency of the external oscillator $\omega_r=0.01148149240963448$ and at the frequency $\omega=0.07$ corresponding to the ground level. The value of the coupling constant between the proton coordinate and that of the oscillator is $\alpha=5$ and the ratio of the mass for the proton to that of the oscillator is $\delta=1$. The value $\beta=0.0345$ corresponds to $T=400 K$ and $\beta=0.069$ to $T=200 K$.} \label{Fig.3}
\end{figure}\\

A descriptive physical origin of the phenomenon of vibrationally enhanced tunneling in our opinion seems to be an analog of the Rabi transition in a two-level system but between the wells of DWP rather than merely between the energy levels. For our symmetric TDWP a direct transition between the energy levels in a doublet within the common well to them is impossible under the influence of a quadratically coupled mode. To see it let us consider a doublet (in our particular case it is the second doublet $n=1$) which is assumed to be responsible for the resonant activation in our system. The wave function in the left well is $\psi_L=1/\sqrt 2\left(\psi_{2n+1}+\psi_{2n}\right)$ and in the right well is $\psi_R=1/\sqrt 2\left(\psi_{2n+1}-\psi_{2n}\right)$ where $\psi_q$ is given by (\ref{eq6}). Here $2n+1$ refers to the upper energy level in the doublet and $2n$ refers to the lower one. The interaction energy term in our case of the symmetric coupling ($\gamma=0$ and $\alpha \not=0$)  is $\sqrt \omega \alpha z\sin^2 x $. A transition between the energy levels in the doublet within the common well to them is forbidden  because the corresponding wave functions are of different parity and the matrix element of the dipole moment is zero
\[
<\psi_{2n+1} \mid \sin^2 x \mid \psi_{2n}>=\int_{-\pi/2}^{\pi/2} dx\ \psi_{2n+1}\sin^2 x\ \psi_{2n}=0
\]
However the transition becomes allowed if we take into account both wells because the matrix element in this case is
\[
<\psi_R \mid \sin^2 x \mid \psi_L>=\int_{-\pi/2}^{\pi/2} dx\ \psi_R\sin^2 x\ \psi_L= \frac{1}{2}\left[\int_{-\pi/2}^{\pi/2} dx\ \psi_{2n+1}^2\sin^2 x-\int_{-\pi/2}^{\pi/2} dx\ \psi_{2n}^2\sin^2 x\right]\not=0
\]
The non-zero value of the matrix element of the dipole moment for the transition between the energy levels in the doublet makes efficient absorbtion of energy from the oscillator to be feasible at a suitable frequency.\\

Mathematically the phenomenon of resonant activation (the high peak in Fig.2) arises from small values of $\tilde\psi_2^2 \left(x_{min}\right)$ at the corresponding resonance frequency $\omega_r$. Unfortunately constructing an empirical rule for the position of the resonance peak is hindered because the dependence of the coefficients $c_{q}(\omega)$ on frequency (which appears in the first order of the adiabatic approximation) makes any equation to be hardly solvable. However we know these coefficients at the resonance frequency from our stringent analysis. We find it expedient to present a simple qualitative way for "deriving" the resonance frequency. The notion of the Rabi frequency is introduced for the dipole approximation of the interaction of some electromagnetic field with energy levels of a quantum system. The Rabi frequency in energetic units (multiplied by the Planck constant) is actually the module of the interaction energy, i.e., that of the product of the field strength and the matrix element of the dipole moment for the transition between the corresponding energy levels. The Rabi resonance condition is $E_f-E_i = \hbar \Omega_{if}^{Rabi}=\mid H_{int} \mid$. By analogy with the Rabi transition we equate the value $\epsilon_{2n+1}-\epsilon_{2n}$ (where $\epsilon_q$ is given by (\ref{eq7})) to the module of the matrix element of the interaction energy term $ \sqrt \omega \alpha z\sin^2 x $ calculated with the functions $\psi_R$ and $\psi_L$ at the corresponding resonant frequency $\omega_r^{(2n)}$ (as mentioned above in our case only the second doublet $n=1$ contributes into PT rate enhancement so that our approximate $\omega_r^{(2)}$ should be identified with $\omega_r$ from the stringent analysis)
\begin{equation}
\label{eq47} \epsilon_{2n+1}-\epsilon_{2n}=\sqrt{\omega_r^{(2n)}}\alpha \mid z <\psi_R \mid \sin^2 x \mid \psi_L>\mid
\end{equation}
The main problem is what we should take for $z$. A reasonable choice is to take the average but there is a duality because we have 
$\bar z_{2n}$ and $\bar z_{2n+1}$. We further write formulas for the case of $\bar z_{2n}$ keeping in mind that the other option may give better agreement. So we change $z$ by its average $\bar z_{2n}$ given by (\ref{eq31})
\begin{equation}
\label{eq48}z\longrightarrow\bar z_{2n}=-\frac{c_{2n}\left(\omega_r^{(2n)}\right)}{\left(\omega_r^{(2n)}\right)^{3/2}}
\end{equation}
It contains the values $c_{2n}\left(\omega_r^{(2n)}\right)$ which as mentioned above we know from our previous analysis. Substituting (\ref{eq48}) into (\ref{eq47}) we have
\begin{equation}
\label{eq49} \epsilon_{2n+1}-\epsilon_{2n}=\alpha  \mid<\psi_R \mid \sin^2 x \mid \psi_L>\mid\frac{c_{2n}\left(\omega_r^{(2n)}\right)}{\omega_r^{(2n)}}
\end{equation}
Restricting ourselves by the second doublet ($n=1$) we take $c_{2n}\left(\omega_r^{(2n)}\right)$ as $c_{2n}\left(\omega_r\right)$.
From here we obtain for the resonance frequency $\omega_r^{(2)}$ 
\begin{equation}
\label{eq50}\omega_r^{(2)}=\alpha  \mid<\psi_R \mid \sin^2 x \mid \psi_L>\mid\frac{c_2\left(\omega_r\right)}{\epsilon_3-\epsilon_2}
\end{equation}
We consider our case $\alpha=5$. In this case we have from the stringent analysis for the second doublet $c_2(\omega_r)\approx0.97$. Calculating the matrix element with the functions $\psi_L=1/\sqrt 2\left(\psi_3+\psi_2\right)$ and $\psi_R=1/\sqrt 2\left(\psi_3-\psi_2\right)$ (where $\psi_q$ is given by  (\ref{eq6})) at the resonant frequency $\omega_r\approx 0.0115$ we obtain from (\ref{eq50}) $\omega_r^{(2)}\approx 0.0126$. For our qualitative argumentation and an extremely rough estimate it is excelent agreement. In our opinion it can not be fortuitous and strongly testifies in favor of the validity of our treating the phenomenon under consideration. Thus we believe that the physical interpretation of PT resonant activation is an analog of the Rabi transition. It takes place under the influence of the vibration with a suitable frequency applied to the proton in DWP. In doing so one should take into account that the transition occurs between the left and the right wells rather than between the energy levels within the common well to them.

A complete parametric analysis of the present model seems to be a daunting task. Here we report only some preliminary results along this line. At our given shape of TDWP (i.e., at the given set of the parameters $\{m=57, p=76\}$) and the given value $\delta=1$ there is a shift of the resonance frequency to lower values with the decrease of $\alpha$. For instance at $\alpha=1$ we obtain $\omega_r\approx0.000459$ while the hight of the peak remains the same. The formula (\ref{eq50}) in this case yields the value $\omega_r^{(2)}\approx 0.0005$ which is in satisfactory agreement. The dependence of the results on  $\delta$ is exremely weak in a very wide range $10^{-2} <\delta<10^{2} $. The alteration of the shape of TDWP changes the picture drastically but retains the existance of the phenomenon of resonant reaction activation. For instance we present the results for $\{m=57, p=70.5\}$ which correspond to the Zundel ion with the distance between oxygen atoms $R_{OO}\approx 2.8\ \AA$. 
\begin{figure}
\begin{center}
\includegraphics* [width=\textwidth] {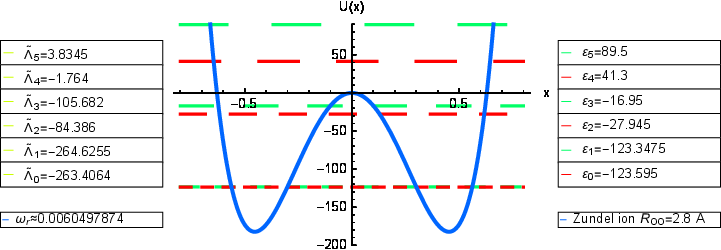}
\end{center}
\caption{The trigonometric double-well potential (\ref{eq2}) at the values of the parameters $m=57$; $p=70.5$. The parameters are chosen to describe the hydrogen bond in the Zundel ion ${\rm{H_5O_2^{+}}}$ (oxonium hydrate) for the case $R_{OO}=2.8\ \AA$ (they are extracted from the data of quantum chemistry \cite{Yu16}, \cite{Xu18}). At the right of Fig.4 the energy levels in the absence of an external oscillation (given by (\ref{eq7})) are presented. The energy levels of polaritonic states (given by (\ref{eq36})) are presented (but not ploted) to the left of Fig.4. They are calculated for the coupling constant $\alpha=5$ and at the frequency of the external oscillator $\omega_r=0.0060497873612747$ corresponding to the peak in Fig.5.} \label{Fig.4}
\end{figure}
In this case there are two doublets below the barrier top (see Fig.4) that means $N=1$ in (\ref{eq46}) (we retain the value $N_{max}=7$ as in the previous case). Fig.5 shows that there is similar effect of PT rate enhancement up to 27 orders of magnitude compared to a non-resonant case. The resonance frequency is $\omega_r\approx 0.00605$ while (\ref{eq50}) in this case yields $\omega_r^{(2)}\approx 0.0042$. Somewhat better matching takes place if (recalling the above mentioned duality) we take in  (\ref{eq50}) the value of $c_3(\omega_r)$ instead of $c_2(\omega_r)$ that yields $\omega_r^{(2)}\approx 0.0055$.  Less satisfactory agreement compared to previous TDWP ($R_{OO}=3.0\ \AA$) may be due to the fact that in this case the second doublet responsible for the phenomenon is rather close to the barrier top. It is well known that at the specified conditions WKB formulas (which are at the basis of the Weiner's theory) become very inaccurate or even invalid \cite{Lan74}.  Further lowering the barrier hight compared to the case $\{m=57, p=70.5\}$  leads to the situation that the second doublet sooner or later becomes over the barrier top.  As namely the second doublet is responsible for the phenomenon of resonant activation we conclude that for low barriers it is hardly possible.
\begin{figure}
\begin{center}
\includegraphics* [width=\textwidth] {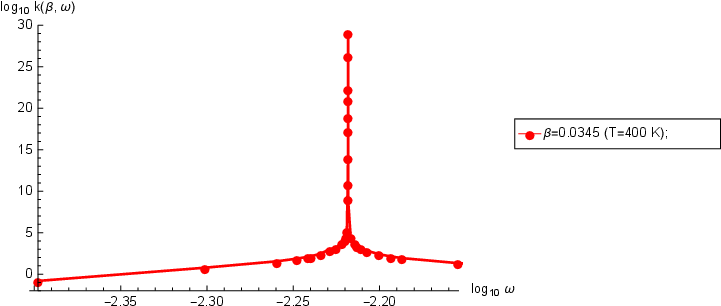}
\end{center}
\caption{The dependence of the proton transfer rate constant on the frequency of the external oscillator in the Zundel ion ${\rm{H_5O_2^{+}}}$ (oxonium hydrate) with $R_{OO}=2.8\ \AA$ (the values of the parameters for TDWP are $m=57$; $p=70.5$) at high temperature $T=400\ K$ ($\beta=0.0345$). The value of the coupling constant between the proton coordinate and that of the oscillator is $\alpha=5$ and the ratio of the mass for the proton to that of the oscillator is $\delta=1$. } \label{Fig.5}
\end{figure}

In the present article we consider VSC as compared to symmetric, anti-symmetric and squeezed mode couplings treated in \cite{Sit23}. There the effect of resonant activation at vibrationally enhanced tunneling was obtained for the case of symmetric mode coupling with no dependence of its strength of interaction on the frequency of the external oscillator. The physical motivation for VSC is discussed in Introduction. Besides we substantially improve the way of calculations compared to \cite{Sit23} in the following aspect. There the two-dimensional SE was solved in the zero-order of the adiabatic approximation. In the present article we carry out the calculations up to the first order. Nevertheless despite these considerable alterations and improvements our calculations for VSC in the symmetric case $\gamma=0$ ($C_1=0$) exhibit the same effect of enormous reaction acceleration in a very narrow frequency range as in  \cite{Sit23}. One can conclude that this phenomenon of resonant activation at vibrationally enhanced tunneling in our model is robust and stable both to variations in the particular kind of the symmetric coupling and in the way of analysis.

Let us briefly touch upon the perspectives of physical implications of the results obtained. The symmetric case $\gamma=0$ ($C_1=0$) considered in the present article is rather restrictive and can not be directly applied to polariton chemistry because light-matter interaction certainly requires the linear term of the dipole moment, i.e. $C_1\not=0$. On the other hand symmetric mode coupling is not of pure academic interest but a quite practical case. For instance it is traditionally used in the theory of HB for the interaction of the heavy atoms stretching mode with the proton coordinate [37-46,65,66,68]. Unfortunately the frequency of this mode is a fixed value that can hardly be varied in an experiment (at least in sufficiently wide range). For this reason such experiments can not be used for verifying our model. The implication  of the results to EHT is a pure speculation at present. No physical model for coupling the reaction coordinate of a corresponding enzyme to a dynamical mode of its portein scaffold is suggested in the present article. Nevertheless the consideration of EHT requires a mathematical model which reliably testifies that the phenomenon of vibrationally enhanced tunneling is a realistic effect. Such model must prove that the mechanism based on this effect is able to provide the resonance activation of PT in HB and yield the desired values of reaction rate enhancement. In our opinion the results obtained are a step towards this aim. In our model we obtain PT acceleration which is sufficient for the necessities of EHT (as mentioned in Introduction the known values of reaction acceleration for EHT are $10^{4}$ and $10^{12}$ at room temperature \cite{Sch09}). By attaining the resonance condition one can obtain an extremely efficient mechanism of PT rate enhancement in our model. Fig.2 and Fig.5 testify that at the resonance frequency one can achieve the reaction acceleration up to $26 \div 27$ orders of magnitude compared to a non-resonant case. Such effect with a safety margin of sufficiency provides the values required for interpretation of the available experimental data in the field of EHT. Morover such reaction acceleration is commensurable with the examples of highest reaction acceleration in enzyme catalysis (see Introduction). Lower reaction acceleration (such as, e.g., $10^{4}$) can be obtained by slight corresponding off-resonance detuning (e.g., in Fig.2 at $\Delta\omega=\mid \omega-\omega_r\mid\approx 10^{-5}$). Finally it is of importance that we obtain the temperature dependence of the rate constant which is a characteristic feature for enzymes carrying out EHT. We make use of the Weiner's theory in which the rate constant is determined by the temperature-averaged flux over all doublets with populations determined by the corresponding Boltzmann factors. The population of the second doublet responsible for the phenomenon is a highly temperature-dependent value that provides the required feature. 

We conclude that our approach enables one to obtain a reliable solution of the two-dimensional Schr\"odinger equation with the Pauli-Fierz Hamiltonian and the trigonometric double-well potential. The calculation is carried out in the first-order of the adiabatic approximation. To be specific we consider the case of vibrational strong coupling which is pertinent for polariton chemistry and (presumably) for enzymatic hydrogen transfer. As an example of applications the solution is used for the calculation of the proton transfer rate constant in the hydrogen bond of the Zundel ion. It is done within the framework of the Weiner's theory. We show that the solution enables one to obtain an analytically tractable expression for the proton transfer rate constant. The parameters of the model for the Zundel ion are extracted from the literature data on IR spectroscopy and quantum chemical calculations. The approach yields a vivid manifestation of the phenomenon of vibrationally enhanced tunneling, i.e., a sharp peak of the reaction rate enhancement by the external vibration at its symmetric coupling to the proton coordinate. The results of the present article along with those of the previous one \cite{Sit23} testify that the effect of resonant activation in our model is sound and stable to variations in the types of the quadratically coupled mode (vibrational strong coupling or symmetric one). Also the existance of the phenomenon proves to be stable to variations in the way of analysis.

\section{Appendix 1}
We consider the dimensional form of stationary SE $\left(H-E\right)\Phi(X,Z)=0$ with PFH (see (1) in \cite{Tri20})
\begin{equation}
\label{eq51}  H=-\frac{\hbar^2}{2M}\frac{\partial^2}{\partial X^2}+V(X)-\frac{\hbar^2}{2\mu}\frac{\partial ^2 }{\partial Z^2}+\frac{\mu}{2}\Omega^2 Z^2+\sqrt {2\Omega} \Upsilon Z Dd\left(X \right)
\end{equation}
We choose the finite interval of the spatial variable $-L \leq X \leq L$ and assume $V(X)$ to be a DWP taking infinite values at its boundaries $X=\pm L$. We identify the normal mode $X$ with that of the proton in HB with the mass $M$. The effective mass of the oscillator with the coordinate $Z$ and the frequency $\Omega$ is assumed to be $\mu$ in the same units as those of $M$. In the vibrational polariton chemistry the oscillator is a mode of the IR electromagnetic field and the dipole moment $D\left(X \right)=Dd\left(X \right)$ interacts with the square-root amplitude of the vacuum fluctuations $\Upsilon$ of the electric component. For EHT the oscillator is a dynamic mode of protein scaffold of an enzyme. Hence $\Upsilon$ in this case is merely a parameter characterizing the coupling of the oscillator to the proton in HB.

The dimensionless values for the coordinates $x$, $z$ and the potential $U(x)$  are introduced as follows
\begin{equation}
\label{eq52} x=\frac{\pi X}{2L};\ \ \ \ \ \ \ \ \ \ z=\frac{\pi Z}{2L};\ \ \ \ \ \ \ \ \ \ U(x)=\frac{8ML^2}{\hbar^2 \pi^2}V(X);\ \ \ \ \ \ \ \ \ \ \epsilon=\frac{8ML^2}{\hbar^2 \pi^2}E
\end{equation}
where $-\pi/2\leq x \leq \pi/2$ and $-\infty < z < \infty$. We introduce the dimensionless parameter of mass ratio $\delta$, the dimensionless frequency $\omega$ and coupling constant $\tilde \alpha$
\begin{equation}
\label{eq53} \delta=\frac{M}{\mu};\ \ \ \ \ \ \ \ \ \ \ \ \omega=\frac{4\sqrt {2} M L^2}{\sqrt {\delta}\hbar \pi^{3/2}}\Omega;\ \ \ \ \ \ \ \ \ \ \ \ \tilde\alpha=\frac{16\Upsilon M^{1/2}\delta^{1/4} L^2 D}{2^{3/4}\hbar^{3/2}\pi^2}
\end{equation}
As a result we obtain the dimensionless PFH (\ref{eq1}). In the case of the trigonometric DWP (\ref{eq2})
the transformation formulas for the parameters $\{m,p\}$ into $\{B,W\}$ ($B$ is the barrier hight and $W$ is the barrier width) are \cite{Sit19}
\begin{equation}
\label{eq54}p=\frac{\sqrt {B}}{1-\left[\cos\left(W/2\right)\right]^2};\ \ \ \ \  \ \ \ \ \ \ \ \ \ \ \ \ \ \ \ \ \  m^2-\frac{1}{4}=\frac{B\left[\cos\left(W/2\right)\right]^4}{\left\{1-\left[\cos\left(W/2\right)\right]^2\right\}^2}
\end{equation}
Introducing the dimensionless energy $\Lambda$ and dipole moment $d(x)$
\begin{equation}
\label{eq55}\Lambda=\frac{8ML^2E}{\hbar^2 \pi^2};\ \ \ \ \ \ \ \ \ \ \ \ \ \ \ \ \ \ \ \ \ \ \ \ \ d(x)=\frac{D\left(2Lx/\pi \right)}{D}
\end{equation}
we obtain the dimensionless form of the two-dimensional SE (\ref{eq4}). The dimensionless inverse temperature $\beta$ in (\ref{eq42})-(\ref{eq44}) is introduced as
\begin{equation}
\label{eq56}\beta=\frac{\hbar^2\pi^2}{8ML^2k_BT}
\end{equation}

The transform to the dimensional rate constant for a particular experiment is carried out with the help of an empirical constant  C. Its choice is determined by fitting experimental data and further for the sake of definiteness we compare a EHT reaction with the corresponding nonenzymatic one. In fact C is related to the shape of TDWP and prior fitting one should choose the barrier hight and width (i.e., the parameters $\{p,m\}$) basing on available data of quantum chemichal calculations or IR spectroscopy for  nonenzymatic case. Then in accordance with the concept discussed in Introduction we assume that the reaction acceleration is not due to barrier hight reduction by the enzyme but is produced by coupling to a dynamical mode (oscillator) in its protein scaffold. It means that we choose $\alpha, \omega_r, ...$ providing the desired reaction acceleration in the dimensioless form (in a specific analog of Fig.2 or Fig.5 from the present paper). The dimensionless rate constant is relatated to the corresponding dimensional one $\Gamma(T, \Omega)$ by the coefficient
\begin{equation}
\label{eq57}
 \frac{\Gamma(T, \Omega)}{k(\beta, \omega)}=C\cdot\frac{\hbar}{ML^2}\sim C\cdot10^{13}\ s^{-1}
\end{equation}
if we take the proton mass $M\sim 10^{-24}g$ and the distance $L\sim 1\ \AA$. Let us consider for example the case of EHT by the alcohol dehydrogenase which provides reaction acceleration $10^{4}$ at room temperature $T\approx 300\ K$  \cite{Sch09}.  In the dimensionless form we choose some reference non-resonant frequency $\omega^{ref}$ on the ground level (plateau) for which the rate constant is $k^{ref}\left(0.057,\omega^{ref}\right)$. As a result the value of the difference $\log_{10}k\Bigl(0.057, \omega_{r}\Bigr)-\log_{10}k^{ref}\left(0.057,\omega^{ref}\right)$ at the resonant frequency $\omega_{r}$ yields the order of reaction acceleration (4 in our example). The rate constant of the corresponding nonenzymatic reaction is $\Gamma^{ref}(T\approx 300\ K)\approx 10^{-3}\ s^{-1}$ \cite{Sch09}. Then the constant $C$ is defined from the relationship
\begin{equation}
\label{eq58}
k^{ref}\left(0.057,\omega^{ref}\right)\cdot C\cdot\frac{\hbar}{ML^2}=\Gamma^{ref}(T\approx 300\ K)
\end{equation}

\section{Appendix 2}
In the Weiner's theory \cite{Wei78}, \cite{Wei78a} the particle position is described by the stationary one-dimensional SE with symmetric DWP $V(X)$ which has the solutions for the energy levels $E_q$ and the corresponding wave functions $\psi_q (X)$
\begin{equation}
\label{eq59}  \frac{\hbar^2}{2M}\psi''_q(X)+\left[E_q-V(X)\right]\psi_q(X)=0
\end{equation}
The rate constant consists of the contribution from the tunneling process and that from the over-barrier transition. Concerning the former the Weiner's theory deals with two important values. The first one is the probability flux to the right of particles in the left well when the particle is in the q-th state $J_q$. The second one is the quantum transmission coefficient, i.e., the fraction of those right-moving particles which are transmitted to the right well $\mid T_q \mid^2$. According to \cite{Wei78}, \cite{Wei78a} the reaction rate constant is a result of Boltzmann averaging of the product $J_q\mid T_q \mid^2$ calculated over the doublets
\begin{equation}
\label{eq60}  k=\left[\sum_{q=0}^{\infty}e^{-E_q/\left(k_BT\right)}\right]^{-1} \left\{\sum_{n=0}^N e^{-E_{2n}/\left(k_BT\right)} J_{2n} \mid T_{2n} \mid^2+\frac{\hbar}{ML^2}\sum_{m=2N+2}^{\infty}e^{-E_m/\left(k_BT\right)}\right\}
\end{equation}
where $n=0,1,2,..., N\ $, $E_{2n}$ is the energy for the level $2n$ described by the wave function $\psi_{2n} (X)$. In the Weiner's theory the quantum transmission coefficient is calculated for the doublets which are counted by the even energy levels. For this reason $n$ is fixed to be even in the first sum in the curly brackets (see the text below the formula (3.1) in Sec.III of \cite{Wei78a} the formulas from which are used in the present article). The first sum in the curly brackets corresponds to the contribution due to the tunneling process in the reaction rate. It is over the energy levels below the barrier top for which the notions of $ J_{2n}$ and $ \mid T_{2n} \mid^2$ have sense. In (\ref{eq60}) it is suggested by Weiner that the quantum transmission coefficient of the lower level in the doublet is determined by the splitting of the energy levels in it. Thus $N+1$ is the number of doublets below the barrier top and $E_{2N}$ is the lower energy level in the last doublet in this region. As a result only the sum over doublets (i.e., even levels $q=2n$) is left. The second sum in the curly brackets corresponds to the over-barrier transition and $E_{2N+2}$ is the first energy level above the barrier top. The Weiner's theory is based on the quasi-classical approximation of the solution of SE \cite{Wei78a}
\begin{equation}
\label{eq61}  \psi_{2n} (x)=\frac{B_{2n}}{\sqrt {P_{2n}(x)}}\cos\frac{1}{\hbar}\left(\int_0^x d\xi\ P_{2n}(\xi) + S_{2n}\right)
\end{equation}
for $x\geq 0$.
The expression for  $\mid T_{2n}\mid^2$ follows from (3.8) and (2.17) of \cite{Wei78a}
\begin{equation}
\label{eq62}  \mid T_{2n} \mid^2=\left[\frac{E_{2n+1}-E_{2n}}{4\hbar J_{2n}}\right]^2
\end{equation}
The expression for $J_{2n}$ is given by (2.14) of \cite{Wei78a}
\begin{equation}
\label{eq63}  J_{2n}=\frac{B_{2n}^2}{4M}
\end{equation}
Substitution of the results into (\ref{eq60}) yields
\begin{equation}
\label{eq64}  k=\left[\sum_{q=0}^{\infty}e^{-E_q/\left(k_BT\right)}\right]^{-1}\left\{\frac{M}{4\hbar^2}\sum_{n=0}^N e^{-E_{2n}/\left(k_BT\right)} \frac{\left[E_{2n+1}-E_{2n}\right]^2}{B_{2n}^2}+\frac{\hbar}{ML^2}\sum_{m=2N+2}^{\infty}e^{-E_m/\left(k_BT\right)}\right\}
\end{equation}

\section{Appendix 3}
Here we present the list of abbreviations used in the article.\\
SE   - Schr\"odinger equation\\
HB   - hydrogen bond\\
PT   - proton transfer\\
EHT  - enzymatic hydrogen transfer\\
DWP  - double-well potential\\
TDWP - trigonometric double-well potential\\
PFH  - Pauli-Fierz Hamiltonian\\
WKB  - Wentzel-Kramers-Brillouin method or quasi-classical approximation\\
VSC  - vibrational strong coupling\\
IR   - infra-red\\
CHF - confluent Heun's function\\
SF -   spheroidal function \\

\section*{Acknowledgments}
The author is grateful to Prof. Yu.F. Zuev and R.Kh. Kurbanov for helpful discussions. The work was supported from the government assignment for FRC Kazan Scientific Center of RAS.

\end{document}